\documentclass{article} 
\usepackage{nips13submit_e,times}
\usepackage{hyperref}
\usepackage{url}
\usepackage{natbib}

\title{Towards User Profile Modelling in Recommender System}

\author{
Djallel Bouneffouf \\
Department of Computer Science\\
Telecom SudParis\\
\texttt{Djallel.bouneffouf@it-sudparis.eu}
}
\begin{document}

\maketitle

\begin{abstract}
The notion of profile appeared in the 1970s decade, which was mainly due to the need to create custom applications that could be adapted to the user. 
In what follows, we treat the different aspects of the user's profile, defining it, profile, its features and its indicators of interest, and then we describe the different approaches of modelling and acquiring the user's interests.   
\end{abstract}

\section{Introduction}

Recommender System (RS) belong to a more general framework of systems called Personalized Access Systems. These systems integrate the user as an information structure, in the process of selecting relevant information to him \cite{Profile}.
\\
In a RS, the user's profile is a kind of query about his interests, describing features that can be shared with a group of individuals. Comparing these features to incoming documents, the system can select those likely to be of interest \cite{baeza1999}.

\subsection{User's Profile Features} 

The features that characterize the user's profile are his acquired knowledge in different subjects (background), his objectives (goals) and his interests \cite{Brusilovsky2011}.
\\
\textbf{Background}: The background concerns all the information related to user's past experience. This includes his profession, experience in fields related to his work, and how the user is familiar with the working environment of the system.
\\
\textbf{Objectives}: The objectives are the user's needs when he searches for information.
\\
\textbf{Interests}: The interests are the documents that the user has consulted or notified to the system directly or indirectly through its feedback by indicators of interests.
\\ 
\subsection{Indicators of Interests}
The implicit detection of interests is through observable behaviours collected by the system when the user interacts with his environment. In this context several behaviours can be considered, such as:
\\
- Click of the mouse on a document;\\
- Scrolling a document using the mouse or keyboard;\\
- Time spent on a document;\\
- Printing a document;\\
- Saving a document;\\
- Copying/pasting all or part of a document;\\
- Document annotation;\\
- Qualitative evaluation of a document;\\
- Navigating to reach a document;\\
- Eye-tracking.\\
These behaviours during user/system interactions are indicators of the relevance of a document and provide useful evidence to predict interests indirectly expressed by the user.
\\
\\
The structures that store the user's profile features are modelled as described in what follows. 

\section{Modelling the User's Profile}
User modelling is a research field that concerns the improvement of man-machine interaction by predicting the interests of users \cite{Profile}.
\\
Modelling the user's profile consists of designing a structure for storing all the information which characterizes the user and describes his interests, his background and his objectives \cite{Salton1975}.
\\
There are several ways to model or represent the user's profile. We describe here the different existing techniques.

\subsection{Vector Representation}
Vector representation is based on the classic vector space model of Salton \cite{Salton1975}, where the profile is represented as an \textit{m}-dimensional vector, where each dimension corresponds to a distinct term and \textit{m} is the total number of terms that exist in the user's profile.
\\
The vector representation has been the first model of the user's profile exploited. The weighting of terms is usually based on a diagram of the TF/IDF format commonly used in IR \cite{Salton1975}. 
\\
The weight associated to each term represents the degree of importance in the user's profile. Different RS use such representation, like \cite{SVM2008} an on line newspaper or \cite{SVM2011} concerning the recommendation of web services.
\\ 
In addition to its simplicity of implementation, the use of several vectors to represent the profile permits to take into account the different interests and their evolution through time; but the default of this representation is in the lacks of structure and semantic (no connexion between terms) .

\subsection{Connexion Representation}
The connexion representation is based on an associative interconnection of the nodes representing the user's profile. 
\\
In this context, the system proposed by \cite{Stef1998} uses the concepts existing in Word-Net to group similar terms. The user's profile is then represented as a semantic network in which each group of concepts is represented by nodes and arcs. 
\\
A similar approach has been used by \cite{Gentiliet2003}. Initially, each semantic network contains a collection of nodes in which each node represents a concept. The nodes contain a single vector of weighted terms. When a new user information is collected, the profile is enriched by integrating weighted terms in the corresponding concepts.
\\
This representation has the advantage of structuring \cite{Stef1998}, but the problem of the connexion representation is the absence of hierarchical relation among the concepts of the network, reflecting the semantic generalization / specialization between concepts (e. g., Biology is one specialization of Science). 

\subsection{Ontologies Representation} 
Ontologies may be used to represent the semantic relations among the informational units that make the user's profile \cite{ontologiesprofile}. This representation allows to overcome the limitations of the connexion representation by presenting the user's profile in the form of a hierarchy of concepts.
\\ 
Each class in the hierarchy represents the knowledge of an area of the user's interests. The relationship (generalization / specification) between the elements of the hierarchy reflects a more realistic interest of the user. 
\\
The representation of the profile based on ontologies creates some problems related to the heterogeneity and diversity of the user's interests, for-instance, the users may have different perceptions of the same concept, which leads to inaccurate representations \cite{godoy2005}.

\subsection{Multidimensional Representation}
The user's profile can contain several types of information such as demographics, interests, purpose, history and other information \cite{multidimensionalprofile}. 
\\
In \cite{multidimensionprofile2011}, the authors represent the contents of the user's profile by a structured model with a  predefined category called dimensions: personal data, data source, data delivery, behavioural data and security data. This model was proposed in the development of a digital library service.
\\
In the same context, authors in \cite{KostadinovBL07} propose a set of open dimensions that can contain most of the information that characterizes the user. The authors propose eight dimensions: personal data, the focus, the domain ontology, the expected quality, customization, security, preferences, miscellaneous information.
\\
The multidimensional representation has the advantage of providing a better interpretation of the semantics of the user's profile.

\section{Acquiring the User's Profile} 
In the recommendation process, the crucial step is the construction of the user's profile that truly reflects his interests. However, this step is not an easy task because the user may not be sure on his interests \cite{dynamicapprochprofile}, on one hand and often does not want or even can not make efforts for its creation, on the other hand. 
\\
In this context, several approaches have been proposed for the acquisition of the user's profile.
\subsection{Simplistic Approach }
This approach promotes the description of the user's profile through a set of keywords explicitly provided by the user \cite{baeza1999}. The method requires much from the user because, if he is not familiar with the system and the vocabulary of incoming documents, it becomes difficult for him to provide the proper keywords that describe his interests. 

\subsection{Dynamic Approach}
An alternative approach is to build the user's profile by dynamically collecting information about his preferences. At the first step, the user must provide a set of keywords describing preferences to initialize the profile; then, at each arrival of a new document, the system uses the user's profile to select the documents potentially fitting his interests, and displays them to the user. 
The user indicates one relevant document and one irrelevant document to him. This information is used to adjust the description of the user's profile in order to reflect his new preferences \cite{baeza1999}.

\subsection{Machine Learning Approach}
Machine learning is an essential step in designing an automatic RS \cite{Lopez2009}. As it is not reasonable to ask the user a set of keywords describing his preferences, the idea is to observe the user's behaviour through his interactions with the system to learn his profile. 
\\
Most systems construct the user's profile by learning from consulted documents. This profile is generally based on the vector model and different indicators of interests are used, such as the movements and mouse clicks, for example. The terms weight adjustment is often used with learning techniques such as neural networks, genetic algorithms and others \cite{Anderson2002}.

\section{Conclusion and Discussion}
We present now a synthesis of the different approaches to represent and acquire the user's profile discussed above.
\begin{table} [h]
\caption{ Representing the user's profile}
\label{tab:Representingfile}       
\begin{tabular}{|p{4cm}|p{5cm}|p{5cm}|p{3cm}|}
\hline
\bf Representation technique    & \bf Advantage & \bf Disadvantage 
 \\
\hline Vector representation & Takes into account the diversity of interests and their evolution through time 
 and is easy to implement & The lack of structure of the data  
 \\
\hline Connectionist representation & Semantic relationships between interests & The absence of hierarchical relations between the network's concepts
\\
\hline Ontology representation    &  Hierarchical relations between the interests of the user's profile
 & Problems related to the heterogeneity and the diversity of the user's interests 
\\
\hline Multidimensional representation & Better interpretation of the semantics of the user's profile
 & Ambiguity in interpreting the roles of each dimension 
\\
\hline
\end{tabular}
\end{table}

Table ~\ref{tab:Representingfile} summarizes the advantages and disadvantages of different approaches on user's profile modelling.
From the table, we note that the connectionist representation solves the shortcomings of the vector representation by establishing relationships between interests in the user's profile. 
Moreover, the ontology representation allows to expand the limits of connexion representation by including a hierarchy between interests. Finally, the multidimensional representation allows a better interpretation of the semantics of the user's profile.

\begin{table} [h]
\caption{ acquisition of the user's profile}
\label{tab:Buildingprofile}       
\begin{tabular}{|p{4cm}|p{5cm}|p{5cm}|p{3cm}|}
\hline
\bf Representation technique    & \bf Advantage & \bf Disadvantage 
 \\
 \hline
Simplistic approach & The system is sure on the user's interests & Hard for the user to provide the appropriate keywords, the profile is static and the system is user-dependent
\\
\hline Dynamic approach & The system is sure on the user's interests, it allows the system to change of the profile & Hard for the user to provide the appropriate keywords and the system is user-dependent
\\
\hline Machine learning approach  & Learning is not restricted to the keywords already entered by the user  & The problem of cold start (no recommendation when the user profile is empty)
\\
\hline
\end{tabular}
\end{table}
Table ~\ref{tab:Buildingprofile} summarizes the advantages and disadvantages of different user's profile acquisitions approaches.
Another observation is that the learning approach and the dynamic approach are complementary, in the sense that the learning approach solves the user dependence found in the dynamic approach, and the dynamic approach solves the cold start problem that exists in the learning approach.
\\
In short, the best approaches to model and acquire the user's profile are respectively a multidimensional approach and a hybrid between the dynamic and the learning approaches.
\\
The representation and acquisition of the user's profile is a part of the recommendation process. 

\bibliographystyle{named}
\bibliography{profile}

\begin{thebibliography}{}

\bibitem[\protect\citeauthoryear{Anderson}{2002}]{Anderson2002}
Corin~Ross Anderson.
\newblock {\em A machine learning approach to web personalization}.
\newblock PhD thesis, 2002.
\newblock AAI3053472.

\bibitem[\protect\citeauthoryear{Baeza-Yates and
  Ribeiro-Neto}{1999}]{baeza1999}
Ricardo~A. Baeza-Yates and Berthier Ribeiro-Neto.
\newblock {\em Modern Information Retrieval}.
\newblock Addison-Wesley Longman Publishing Co., Inc., Boston, MA, USA, 1999.

\bibitem[\protect\citeauthoryear{Bila \bgroup \em et al.\egroup
  }{2008}]{dynamicapprochprofile}
Nilton Bila, Jin Cao, Robert Dinoff, Tin~Kam Ho, Richard Hull, Bharat Kumar,
  and Paulo Santos.
\newblock Mobile user profile acquisition through network observables and
  explicit user queries.
\newblock In {\em Proceedings of the The Ninth International Conference on
  Mobile Data Management}, MDM '08, pages 98--107, Washington, DC, USA, 2008.
  IEEE Computer Society.

\bibitem[\protect\citeauthoryear{Brusilovsky}{2001}]{Brusilovsky2011}
Peter Brusilovsky.
\newblock Adaptive hypermedia.
\newblock {\em User Modeling and User-Adapted Interaction}, 11(1-2):87--110,
  March 2001.

\bibitem[\protect\citeauthoryear{Chan \bgroup \em et al.\egroup
  }{2011}]{SVM2011}
Nguyen~Ngoc Chan, Walid Gaaloul, and Samir Tata.
\newblock A web service recommender system using vector space model and latent
  semantic indexing.
\newblock In {\em Proceedings of the 2011 IEEE International Conference on
  Advanced Information Networking and Applications}, AINA '11, pages 602--609,
  Washington, DC, USA, 2011. IEEE Computer Society.

\bibitem[\protect\citeauthoryear{Godoy and Amandi}{2005}]{godoy2005}
Daniela Godoy and Anal\'{\i}a Amandi.
\newblock User profiling in personal information agents: a survey.
\newblock {\em Knowledge Eng. Review}, 20(4):329--361, 2005.

\bibitem[\protect\citeauthoryear{Kostadinov \bgroup \em et al.\egroup
  }{2007}]{KostadinovBL07}
Dimitre Kostadinov, Mokrane Bouzeghoub, and St{\'e}phane Lopes.
\newblock Query rewriting based on user's profile knowledge.
\newblock In {\em BDA}, 2007.

\bibitem[\protect\citeauthoryear{Lakiotaki \bgroup \em et al.\egroup
  }{2011}]{multidimensionprofile2011}
Kleanthi Lakiotaki, Nikolaos~F. Matsatsinis, and Alexis Tsoukias.
\newblock Multicriteria user modeling in recommender systems.
\newblock {\em IEEE Intelligent Systems}, 26(2):64--76, 2011.

\bibitem[\protect\citeauthoryear{Lopez-Lopez \bgroup \em et al.\egroup
  }{2009}]{Lopez2009}
L.~M. Lopez-Lopez, J.~J. Castro-Schez, D.~Vallejo-Fernandez, and J.~Albusac.
\newblock A recommender system based on a machine learning algorithm for b2c
  portals.
\newblock In {\em Proceedings of the 2009 IEEE/WIC/ACM International Joint
  Conference on Web Intelligence and Intelligent Agent Technology - Volume 01},
  WI-IAT '09, pages 524--531, Washington, DC, USA, 2009. IEEE Computer Society.

\bibitem[\protect\citeauthoryear{Mezghani \bgroup \em et al.\egroup
  }{2012}]{Gentiliet2003}
Manel Mezghani, Corinne~Amel Zayani, Ikram Amous, and Faiez Gargouri.
\newblock A user profile modelling using social annotations: a survey.
\newblock In {\em Proceedings of the 21st international conference companion on
  World Wide Web}, WWW '12 Companion, pages 969--976, New York, NY, USA, 2012.
  ACM.

\bibitem[\protect\citeauthoryear{Middleton \bgroup \em et al.\egroup
  }{2004}]{ontologiesprofile}
Stuart~E. Middleton, Nigel~R. Shadbolt, and David~C. De~Roure.
\newblock Ontological user profiling in recommender systems.
\newblock {\em ACM Trans. Inf. Syst.}, 22(1):54--88, January 2004.

\bibitem[\protect\citeauthoryear{Mukhopadhyay \bgroup \em et al.\egroup
  }{2008}]{SVM2008}
Debajyoti Mukhopadhyay, Ruma Dutta, Anirban Kundu, and Rana Dattagupta.
\newblock A product recommendation system using vector space model and
  association rule.
\newblock In {\em Proceedings of the 2008 International Conference on
  Information Technology}, ICIT '08, pages 279--282, Washington, DC, USA, 2008.
  IEEE Computer Society.

\bibitem[\protect\citeauthoryear{Nieder\'{e}e \bgroup \em et al.\egroup
  }{2004}]{multidimensionalprofile}
Claudia Nieder\'{e}e, Avar\'{e} Stewart, Bhaskar Mehta, and M.~Hemmje.
\newblock {A Multi-Dimensional, Unified User Model for Cross-System
  Personalization}.
\newblock In {\em E4PIA Workshop 2004}, 2004.

\bibitem[\protect\citeauthoryear{Salton \bgroup \em et al.\egroup
  }{1975}]{Salton1975}
G.~Salton, A.~Wong, and C.~S. Yang.
\newblock A vector space model for automatic indexing.
\newblock {\em Commun. ACM}, 18(11):613--620, November 1975.

\bibitem[\protect\citeauthoryear{Shani \bgroup \em et al.\egroup
  }{2007}]{Profile}
Guy Shani, Lior Rokach, Amnon Meisles, Lihi Naamani, Nischal Piratla, and David
  Ben-shimon.
\newblock Establishing user profiles in the mediascout recommender system,
  2007.

\bibitem[\protect\citeauthoryear{Wibowo \bgroup \em et al.\egroup
  }{2011}]{Stef1998}
Adi Wibowo, Andreas Handojo, and Albert Halim.
\newblock {Application of Topic Based Vector Space Model with WordNet}.
\newblock In {\em International Conference on Uncertainty Reasoning and
  Knowledge Engineering}, 2011.

\end{thebibliography}

\end{document}